\documentclass[epj]{svjour}

\usepackage{graphics}
\usepackage{epsfig}
 \usepackage{graphicx}
 \usepackage{color}
 \usepackage{mathrsfs} 
 \usepackage{lscape,epsfig}
 \usepackage{amsmath}
 \usepackage{bm}
 \usepackage{txfonts}
 \usepackage{multirow}

\usepackage{graphicx}
\usepackage{dcolumn}
\usepackage{bm}
\usepackage{color}
\usepackage{float} 
\usepackage[section]{placeins}

\usepackage{amsmath}
\usepackage{amsfonts}
\usepackage{braket}

\usepackage{fancyhdr}
\usepackage{hyperref}
\def\beq{\begin{equation}}
\def\eeq{\end{equation}}
\def\beqa{\begin{eqnarray}}
\def\eeqa{\end{eqnarray}}

\begin{document}

\title{Neutron Star Properties and Femtoscopic Constraints}

\author{I. Vida\~na\inst{1} \and V. Mantovani Sarti\inst{2} \and J. Haidenbauer\inst{3} \and D. L. Mihaylov\inst{2,4} \and L. Fabbietti\inst{2}}
\institute{
Istituto Nazionale di Fisica Nucleare, Sezione di Catania, Dipartimento di Fisica ``Ettore Majorana'', Universit\`a di Catania, Via Santa Sofia 64, I-95123 Catania, Italy \and
Physik Department E62, Technische Universit\"{a}t M\"{u}nchen, Garching, Germany, EU \and
Institute for Advanced Simulation (IAS-4), Forschungszentrum J\"{u}lich, D-52425 J\"{u}lich, Germany \and
Sofia University, Faculty of Physics, 5 J. Bourchier Blvd., Sofia, Bulgaria
}

\date{Received: date / Revised version: date}


\abstract{We construct the equation of state of hypernuclear matter and study the structure of neutron stars employing a chiral hyperon-nucleon interaction of the J\"{u}lich--Bonn group tuned to femtoscopic $\Lambda p$ data of the ALICE Collaboration, and $\Lambda\Lambda$ and $\Xi$N interactions determined from lattice QCD calculations by the HAL QCD Collaboration that reproduce the femtoscopic 
$\Lambda\Lambda$ and $\Xi^-p$ data. We employ the ab-initio microscopic Brueckner--Hartree--Fock theory extended to the strange baryon sector. A special focus is put on
the uncertainties of the hyperon interactions and how they are effectively propagated
to the composition, equation of state, mass-radius relation and tidal deformability of neutron stars. To such end, we consider the uncertainty due to the experimental error of the femtoscopic $\Lambda p$ data used to fix the chiral hyperon-nucleon interaction and the theoretical uncertainty, estimated from the residual cut-off dependence of this interaction. 
We find that the final maximum mass of a neutron star with hyperons is in the range $1.3-1.4$ $M_\odot$, in agreement with previous works. The hyperon puzzle, therefore, remains still an open issue if only two-body hyperon-nucleon and hyperon-hyperon interactions are considered. {Predictions for the tidal deformability of neutron stars with hyperons are found to be in agreement with the observational constraints from the gravitational wave event GW170817 in the mass range $1.1-1.3$ $M_\odot$.}
}


\maketitle

\section{Introduction}
\label{sec:introduction}

The presence of hyperons in finite and infinite nuclear systems allows us the study of baryon interactions from an enlarged perspective and, consequently, to extend our present knowledge of conventional nuclear physics to the SU(3)-flavor sector \cite{Botta12,Gal16,Lenske18}. Particularly, a big interest has being put 
on the study of hyperonic matter (nuclear matter with nucleonic and hyperonic degrees of freedom) in connection with the physics of neutron star interiors \cite{Compstar}. Neutron star properties such as for instance their masses, thermal evolution, neutrino emissivity, or their gravitational instabilities are closely related to the underlying equation of state (EoS) of matter at high densities \cite{Baldo99}. Therefore, despite the fact that hypernuclear matter is an idealized system, the theoretical determination of its EoS is an essential step towards the understanding of those properties, which can be affected by the presence of hyperons \cite{Chatterjee16,Vidana18}. However, this is a very challenging task since it requires the knowledge of two-, three- and, eventually, many-body interactions involving hyperons and nucleons, and the solution of the very complicated nuclear many-body 
problem \cite{RingSchuck}.

Compared to the nucleon-nucleon (NN) interaction, the hyperon-nucleon (YN) and hyperon-hyperon (YY) ones are still poorly constrained, mainly due to the scarce amount of YN scattering data, and to the complete lack of them in the YY case. The reason are the experimental difficulties associated with the short lifetime of hyperons and the low-beam intensity fluxes. Three-baryon forces involving hyperons (YNN, YYN and YYY)
~\cite{Petschauer:2015elq,Haidenbauer:2016vfq,Kohno:2018gby,Logoteta:2019utx,Gerstung:2020ktv,Tong:2024egi} are even less constrained experimentally, although their role could be fundamental for the solution of the so-called hyperon puzzle, {\it i.e.}, the problem of the strong softening of the dense matter EoS due to the presence of hyperons which leads to maximum masses of neutron stars not compatible with the 
recent observations of $\sim 2 M_\odot$ millisecond pulsars \cite{Demorest10,Antoniadis13,Cromartie20}.  However, information on the two-body
YN interactions, additional to that from scattering experiments and hypernuclei \cite{Botta12,Gal16}, together with constraints on the YY ones and three-body forces has emerged recently thanks to the so-called femtoscopy technique by measuring the correlations (in momentum space) of YN pairs~\cite{ALICE:pXiNature,ALICE:pLcoupled}, YY pairs~\cite{ALICE:LL} and YNN triads \cite{Acharya23} produced in proton-proton (pp) collisions at the LHC by the ALICE Collaboration. The comparison of the measured two- and three-body correlation functions with theoretical predictions permits one to test, improve and constrain the existing two- and three-body hyperon interactions.

The aim of the present work is to construct the EoS of hypernuclear matter and to study the structure of neutron 
stars employing the chiral $\Lambda$N-$\Sigma$N interaction recently tuned to femtoscopic data of the ALICE Collaboration~\cite{Acharya23,Mihaylov24}, 
and $\Lambda\Lambda$ and $\Xi$N interactions determined from the lattice QCD simulations of the HAL QCD Collaboration \cite{Sasaki20} which predict results 
in agreement with femtoscopy. 
In particular, we use the chiral YN interaction NLO19 of the J\"{u}lich--Bonn 
group \cite{Haidenbauer00}, and variants of it recently established in a combined analysis of $\Lambda p$ scattering and correlation-function data
by Mihaylov {\it et al.}~\cite{Mihaylov24}. 
Regarding the $\Lambda\Lambda$ and $\Xi$N interactions we consider two 
specific sets of the HAL QCD potential that were employed in the analysis of 
data on the $\Xi^-p$ 
correlation function by the ALICE Collaboration~\cite{ALICE:pXiNature}. 
The EoS of hypernuclear matter is then constructed by using these interactions within
the well known ab-initio microscopic Brueckner--Hartree--Fock (BHF) theory extended to the strange baryon sector \cite{Schulze95,Schulze98,Baldo98,Baldo00,Vidana00,Vidana00b,Schulze06,Schulze11}. The major novelty of this work is a detailed
study of how the uncertainties on the YN and YY interactions that describe available current experimental data are effectively propagated to the EoS and to the structure of a system as complex as a neutron star with strangeness content. 

The paper is organized in the following way. Details on the YN and YY employed in this work are given in Sect.\ \ref{sec:twobodypot}. The theoretical (BHF) description of the hyperonic matter of the EoS is briefly reviewed in Sect.\ \ref{sec:BHF}. Results for the composition, the EoS and the structure of neutron stars are presented in Sect.\ \ref{sec:results} with a special focus put on the uncertainties of the interactions and how they are effectively propagated to the EoS and the neutron star structure. Finally, a short summary and the main conclusions of this work are given in Sect.\ \ref{sec:conclusions}.

\section{YN and YY interactions}
\label{sec:twobodypot}

In this work we focus on the strangeness $S=-1, -2$ baryon-baryon interactions, considering explicitly the $\Lambda$N-$\Sigma$N and $\Lambda\Lambda$-$\Xi$N coupled channels, respectively.  The novelty of this work lies in the fact that the interactions employed as input to determine the behavior of the hypernuclear
matter EoS yield results in agreement with the recent $\Lambda p$~\cite{ALICE:pLcoupled,Mihaylov:CECA}, $\Lambda\Lambda$~\cite{ALICE:LL}
and $\Xi^-p$~\cite{ALICE:pXiNature,ALICE:pXiPRL} correlation measurements performed by the ALICE Collaboration.  

The $S=-1$ YN interaction is described by the NLO19 potentials derived within SU(3) 
chiral effective field theory ($\chi$EFT) in Ref.\ \cite{Haidenbauer00}. 
These potentials incorporate contributions up to 
next-to-leading order (NLO) 
in the chiral expansion, in particular they include contributions from one- and 
two-pseudoscalar-meson exchange diagrams, involving the
Goldstone bosons $\pi$, $K$, $\eta$, 
and from four-baryon contact terms 
(without and with two derivatives), where the latter
encode the unresolved short-distance dynamics.
The low energy constants (LECs) associated with these contact
terms are free parameters, which need to be constrained by experimental data, and they have been so far established by a global fit to a set of
$36$ $\Lambda p$ and $\Sigma$N low energy scattering data points \cite{Haidenbauer00}, available since the 1960s. 
SU(3) flavor symmetry has been imposed which reduces the number of independent contact terms or LECs, respectively \cite{Haidenbauer:NLO13}. 


Of specific interest for the present study are variants of those potentials which have been established in Ref.~\cite{Mihaylov24}, in the combined analysis of the above mentioned scattering data and the more recent $\Lambda p$ correlation function measured by ALICE \cite{Mihaylov:CECA}. In order to generate a slightly weaker $\Lambda$N interaction, suggested by those correlation data \cite{Mihaylov24}, the possible SU(3) symmetry breaking in the leading-order contact terms of the relevant $^1S_0$ and $^3S_1$ partial waves has been exploited, 
in line with $\chi$EFT and the associated power counting \cite{Haidenbauer:NLO13,Petschauer:2013}, so that the description of the cross sections in other YN reactions ($\Sigma^+ p$, $\Sigma^- p$, $\Sigma^- p$ $\to$ $\Lambda n$) remains practically unchanged as compared to the original potential~\cite{Haidenbauer00}. The allowed 1$\sigma$ region in the singlet and triplet scattering length plane $(f_0,f_1)$ obtained in~\cite{Mihaylov24} delivers an estimate on the current uncertainty arising from all the available experimental input. 
{Besides the uncertainty related to the statistical errors of the ALICE data there is also a theoretical uncertainty.
The main source for the latter is the truncation in the chiral expansion. In 
the present study both sources of uncertainties are estimated. 

Following Ref.~\cite{Mihaylov24}, the data driven errors are considered by employing variants of the NLO19 YN potential with cutoff 600~MeV \cite{Haidenbauer00} whose LECs were readjusted in such a way that the 
singlet ($^1S_0$) scattering length $f_0 = 2.50~$fm is produced and variations of the triplet ($^3S_1$) scattering length $f_1$ between the allowed 
values of $1.32$~fm and $1.55~$fm are achieved, 
see Tab.\ \ref{tab:LN}.
An essential feature of $\chi$EFT is that, contrary to calculations with phenomenological models, an estimation for the theoretical uncertainty can be provided~\cite{Epelbaum:2008ga,Epelbaum:2014efa}. In the present work it is estimated from the dependence of the results on the regulator cutoff. For that, the LECs are readjusted so that for all cutoff values considered in the original NLO19 potential (500-650~MeV) the scattering lengths of NLO19 (600) are obtained \cite{Haidenbauer00}, see variants II in Tab.\ \ref{tab:LN}.
Note that the cutoff dependence provides only a lower bound for the uncertainty. For more elaborated ways to estimate the theoretical uncertainty in the context of nuclear matter calculations, that do not rely on cutoff variation, see~\cite{Hu:2016nkw}. 
}

\begin{table}
{
    \centering
    \renewcommand{\arraystretch}{1.4}
\vskip 0.3cm
\begin{center}
\begin{tabular}{|ll|cccc|}
\hline
           & & $f_0$ & $d_0$ & $f_1$ & $d_1$  \\
\hline \hline
NLO19(600) & &  2.91 & 2.78 & 1.41 &  2.52\\
\hline
variants I &$n_\sigma=-1$         &  2.50 & 2.95 & 1.32 &  2.63\\
&$n_\sigma=0$   &  2.50 & 2.95 & 1.46 &  2.47 \\
&$n_\sigma=1$    &  2.50 & 2.95 & 1.55 &  2.37 \\
\hline
variants II&(500)   &  2.91 & 3.10 & 1.41 &  2.74  \\
& (550)   &  2.91 & 2.93 & 1.41 &  2.66 \\
&(650)   &  2.91 & 2.65 & 1.41 &  2.59 \\
\hline
\end{tabular}
\renewcommand{\arraystretch}{1.0}
\caption{Singlet and triplet $\Lambda N$ scattering lengths ($f_0,f_1$) and effective range ($d_0,d_1$) parameters (in fm)
for the YN interaction NLO19(600)~\cite{Haidenbauer00}
and for the variants~\cite{Mihaylov24} utilized in the
present work. Set I of variants reflects the uncertainty due to the
$\Lambda p$ data (cross sections, correlation functions) while set II
is used to estimate the theoretical uncertainty, based on the residual
cutoff dependence.}
   \label{tab:LN}
\end{center}
}
\end{table}


\begin{table*}
{
\renewcommand{\arraystretch}{1.3}
\begin{center}
\begin{tabular}{|c|c|c|c|}
\hline
 & & $f_0$ [fm]  & $d_0$ [fm] \\
\hline \hline
\multirow{3}{*}{ $^1S_0$ }  & $p\Xi^- $ & 
$ 1.25(0.03)( _{-0.12} ^{+0.00})+i2.00(0.40)( _{-0.16} ^{+0.31})$  & 
$3.7(0.3)( ^{+0.0} _{-0.1})-i2.4(0.2)( ^{+0.1} _{-0.3})$\\
           & $n\Xi^0 $                 
           & $2.76(0.63)( _{-0.33} ^{+0.66})+i 0.15(0.12)( _{-0.00} ^{+0.03})$ &$ 1.5(0.3)( ^{+0.0} _{-0.1})-i0.1(0.0)( ^{+0.0} _{-0.0})$\\
           & $\Lambda\Lambda$ & $0.99(0.30)( _{-0.00} ^{+0.17})$ 
           &$4.9(0.70)( ^{+0.1} _{-0.5})$ \\ \hline
            & & $f_1$ [fm]  & $d_1$ [fm] \\ \hline
 \multirow{2}{*}{ $^3S_1$ }  & $p\Xi^- $                  
 & $ 0.47(0.08)( _{-0.11} ^{+0.09})+i0.0(0.00)( _{-0.00} ^{+0.00})$         
 &$ 6.7(0.7)( ^{+1.4} _{-0.9})+i 0.0(0.1)( ^{+0.0} _{-0.0})$\\
                   & $n\Xi^0 $                 
                   &$ 0.47(0.08)( _{-0.11}^{+0.09})       $ 
                   &$ 6.8(0.7)( ^{+1.4}_{-0.9})$\\
\hline
\end{tabular}
\renewcommand{\arraystretch}{1.0}
\caption{HAL QCD results for the $\Lambda\Lambda$ and $\Xi N$ singlet and triplet scattering length and effective
range parameters in the particle
basis and without Coulomb interaction, taken from Ref.~\cite{Kamiya:2021hdb}.}
\label{tab:aXH}
\end{center}
}
\end{table*}

For the interaction in the $S=-2$ sector we employ the  {$\Lambda\Lambda$, $\Xi$N and $\Lambda\Lambda$-$\Xi$N} {S-wave interactions} obtained by the HAL QCD Collaboration~\cite{Sasaki20}. {A possible coupling to the $\Lambda\Sigma$ and $\Sigma\Sigma$ channels was ignored in that work and is likewise not considered in the present study. The $\Lambda\Lambda$, $\Xi$N and $\Lambda\Lambda$-$\Xi$N} {potentials} were {determined on the basis} of (2+1)-flavor lattice QCD simulations close to the physical point {($m_\pi\approx 146$ MeV and $m_K\approx 525$ MeV), and parametrized as a combination of Gaussian and Yukawa functions. The statistical errors of the fitting parameters of these interactions were estimated considering
three different values of $t/a = 11,12,13$\footnote{Here, $t$ is the Euclidean time and $a$ represents the lattice spacing used in the simulated grid.}
and using the jackknife method \cite{Quenouille49,Tukey58,Miller74,EfronGong1983}, a popular resampling technique of statistics used to calculate the bias and standard error of a statistical estimate. The basic idea behind this method lies in recomputing systematically the estimate a large number of times from subsamples of the available sample, leaving out one element or a group of elements at a time. For a detailed description of the jackknife method the interested reader is referred to Refs.\ \cite{Quenouille49,Tukey58,Miller74,EfronGong1983}.

The $\Lambda\Lambda$ interaction from the lattice simulations shows attraction 
at low energies, though it is not strong enough to generate a bound or resonant dihyperon state around the $\Lambda\Lambda$ threshold. On the other hand, 
the $\Xi$N interaction produces a relatively strong attraction in the spin ($S$) isospin ($I$) channel $(S=0, I=0)$ while it is weakly repulsive in the $(S=0,I=1)$ one, and weakly attractive in the $(S=1, I=0)$ and the $(S=1, I=1)$ ones. 
Regarding the $\Lambda\Lambda$ interaction the scattering length and the effective range predicted by the HAL QCD potential was found to be consistent with the experimental constraints inferred from the analysis of the experimental $\Lambda\Lambda$ correlation measured by the ALICE Collaboration \cite{ALICE:LL}. 
The $\Xi$N interaction deduced from the lattice simulations was used 
directly by the ALICE Collaboration in its analysis of the $\Xi^- p$ correlation function, measured in two different systems, namely in pp \cite{ALICE:pXiNature} and p-Pb \cite{ALICE:pXiPRL} collisions. A good agreement between the correlation measurement in pp collisions and the prediction based on 
the $\Xi$N interaction from the lattice simulation was found. 
In fact, in the analysis by ALICE all $\Xi N$ potentials generated by the
HAL QCD Collaboration in the course of their jackknife evaluation were tested,
in total 71 different parameterizations of the lattice potential \cite{Sasaki:2020}, 
corresponding to the three different $t/a = 11,12,13$ values. 
In the present work, we have selected two specific parameterizations out of these 71 which correspond, respectively, to the most and the least attractive cases for the $\Xi$N interaction compatible with the femtoscopic $\Xi^-p$ data. We note that a detailed analysis of the $\Lambda\Lambda$ and $\Xi^-p$ correlation functions based on the HAL QCD potentials has been also performed by Kamiya {\it et al.}~\cite{Kamiya:2021hdb}, which confirmed the agreement between the theory predictions and experiment. In Tab.~\ref{tab:aXH} we provide an overview of the singlet and triplet scattering length and the effective range parameters predicted by those potentials \cite{Kamiya:2021hdb}.
}

Finally, let us come to the pure nucleonic interaction included in the determination of the EoS. For the two-body NN part, we employ the well-known Argonne V18 potential~\cite{Wiringa95}. Effects of three-nucleon forces (3NFs) are accounted for too, as explained in more detail in the next section.

\section{BHF approach of Hyperonic Matter}
\label{sec:BHF}

\begin{table*}[t!]
\begin{center}
\renewcommand{\arraystretch}{1.3}
\begin{tabular}{|cccc|}
\hline
$a$ (MeV fm$^3$) & $b$ (MeV fm$^6$) & $E_0$ (MeV) & $K_0$ (MeV)   \\
\hline \hline
 $-15.22$ & $21.70$ & $-16$ & $160$   \\ 
 $-53.42$ & $260.42$ & $-16$ & $270$  \\
\hline
\end{tabular}
\end{center}
\renewcommand{\arraystretch}{1.0}
\caption{Values of the coefficients $a$ and $b$ for the two extreme nucleonic EoS considered. The corresponding values of the binding energy $E_0$ and the nuclear incompressibility $K_0$ at saturation density are also given.}
\label{tab:tab1}
\end{table*}
Our description of the hyperonic matter EoS starts with the construction of all the $G$-matrices that describe the interaction of two baryons immersed in a surrounding medium. They are obtained by solving the well known Bethe--Goldstone equation, written schematically as
\begin{eqnarray}
G(\omega)_{B_1B_2;B_3B_4}=
V_{B_1B_2;B_3B_4} \nonumber \\
+\frac{1}{\Omega}\sum_{B_iB_j}V_{B_1B_2;B_iB_j} 
\frac{Q_{B_iB_j}}{\omega-E_{B_i}-E_{B_j}+i\eta}
\nonumber \\
\times G(\omega)_{B_iB_j;B_3B_4}
\label{eq:bbg}
\end{eqnarray} 
where the first (last) two subindices indicate the initial (final) two-baryon state compatible with a given value $S$ of the strangeness,
namely $S=0$ (NN), $S=-1$ ($\Lambda$N, $\Sigma$N) and $S=-2$ 
($\Lambda\Lambda$, $\Xi$N); $V_{B_1B_2;B_3B4}$ is the bare baryon-baryon interaction ($B_1B_2\rightarrow B_3B_4$); $\Omega$ is the (large) volume enclosing the system; $Q_{B_iB_j}$ is the Pauli operator that takes into account the effect of the exclusion principle on the scattered baryons; and the starting energy $\omega$ corresponds to the sum of the nonrelativistic single-particle energies of the interacting baryons. We note that Eq.~(\ref{eq:bbg}) is a coupled channel equation 
(see for instance Ref.\ \cite{Vidana00} for a detailed discussion on the Bethe--Goldstone equation in coupled channels).

The single-particle energy of a baryon $B_i$ with momentum $\vec k$ is given by
\begin{equation}
E_{B_i}(\vec k)=M_{B_i}+\frac{\hbar^2k^2}{2M_{B_i}}+Re[U_{B_i}(\vec k)] \ .
\label{eq:spe}
\end{equation}

Here $M_{B_i}$ denotes the rest mass of the baryon and the single-particle potential $U_{B_i}(\vec k)$ represents the average mean field  ``felt'' by the baryon owing to its interaction with the other baryons of the (hyper)nuclear medium.
In the BHF theory $U_{B_i}(\vec k)$ is calculated through the ``on-shell'' $G$-matrix given by
\begin{eqnarray}
U_{B_i}(\vec k)&=&\frac{1}{\Omega}\sum_{B_j}\sum_{\vec k'}n_{B_j}(|\vec k'|) \nonumber \\
&\times&\langle \vec k\vec k' |G_{B_iB_j;B_iB_j}(\omega)|\vec k\vec k'\rangle_{\cal A}
\label{eq:spp}
\end{eqnarray}
where $n_{B_j}(|\vec k|)=\theta (k_{F_{B_j}} -|\vec k|)$ is the occupation number of the baryon species $B_j$ with Fermi momentum
$k_{F_{B_j}}$, the index ${\cal A}$ indicates that the matrix elements are properly antisymmetrized when baryons $B_i$ and
$B_j$ belong to the same isomultiplet, and $\omega=E_{B_i}(\vec k)+E_{B_j}(\vec k')$. We note here that in this work we have adopted the so-called continuous prescription when solving the Bethe--Goldstone equation, since as shown by the authors of Refs.\ \cite{Song98,Song00} the contribution to the energy per particle of nuclear matter from three-hole line diagrams (which account from three-body correlations) is minimized when this prescription is considered. Once a self-consistent solution of Eqs.\ (\ref{eq:bbg}-\ref{eq:spp}) is achieved, the energy density of the system can be calculated in the BHF theory as
\begin{eqnarray}
\varepsilon_{BHF}&=&\frac{1}{\Omega}\sum_{B_i}\sum_{\vec k}n_{{B_i}}(|\vec k|) \nonumber \\
&\times&\left(M_{B_i}+\frac{\hbar^2k^2}{2M_{B_i}}+\frac{1}{2}U_{B_i}(\vec k)
\right) \ .
\label{eq:edens}
\end{eqnarray}

The inclusion of genuine 3NFs in the microscopic BHF approach is currently a task still far to be achieved because it would require the solution of a three-body 
Bethe--Faddeev equation in the nuclear medium. Therefore, the usual way of introducing them consist in adding an effective density-dependent two-nucleon force to the bare NN interaction when solving the 
Bethe--Goldstone equation. This effective force is obtained by averaging the genuine 3NF over the spatial, spin and isospin coordinates of one of the three nucleons (see {\it e.g.,} Refs.\ \cite{Loiseau71,Grange76,Baldo99b}). In this work, however, we have adopted a simpler alternative strategy which will allow us to explore the uncertainty on the
neutron star properties associated to the softness/stiffness of the pure nucleonic part of the EoS. To such end, first, we construct, as described above, the energy density of hyperonic matter within the BHF approach using only two-body NN, YN and YY forces and then, we add to it the following phenomenological density- and isospin-dependent term that accounts for the effect of 3NFs
\begin{equation}
\varepsilon_{3NF}=(a\rho_N^2+b\rho_N^3)(1+\beta^2) \ .
\label{eq:ct}
\end{equation}
Here $\rho_N=\rho_n+\rho_p$ is the total nucleonic density and $\beta=(\rho_n-\rho_p)/(\rho_n+\rho_p)$ the isospin asymmetry parameter. The coefficients $a$ and $b$ are fitted to reproduce simultaneously the binding energy $E_0=-16$ MeV of symmetric matter at nuclear saturation density $\rho_0=0.17$~fm$^{-3}$ and the nuclear incompressibility $K_0$, for which two extreme values compatible with current experimental uncertainties \cite{Fevre16} have been considered: $K_0=160$ MeV and $K_0=270$ MeV corresponding, respectively, to a very soft and to a very stiff nucleonic EoS. The values of the coefficients $a$ and $b$ are given in Tab.\ \ref{tab:tab1}. 

Once the total energy density $\varepsilon=\varepsilon_{BHF}+\varepsilon_{3NF}+\varepsilon_L$ ($\varepsilon_L$ being the contribution of non interacting leptons) is
known, the composition and the EoS of neutron star matter can be obtained from the requirement of  equilibrium under weak $\beta$ decay processes, $\mu_i=b_i\mu_n-q_i\mu_e$ ($b_i$ and
$q_i$ denoting the baryon number and charge of species $i$)
and electric charge neutrality, $\sum_iq_i\rho_i=0$. The chemical potentials of the various species and the pressure
are computed from the usual thermodynamical relations, $\mu_i=\partial \varepsilon /\partial \rho_i$ and $P=\rho^2\partial (\varepsilon/\rho)/\partial \rho$. 
{To describe the different regions of the stellar crust we use the well-known EoSs of Feynman--Metropolis--Teller \cite{Feynman49}, Baym--Pethick--Sutherland
\cite{Baym71} and Negele--Vautherin \cite{Negele73}.}
Finally, knowing the EoS, the equilibrium configurations of static non-rotating neutron stars are obtained by solving the well-known Tolman--Oppenheimer--Volkoff (TOV) equations \cite{Tolman39,Oppenheimer39}. {To determine the  tidal deformability $\lambda$ of the different star configurations we solve, together with the TOV equations, a first-order ordinary differential equation to calculate the so-called tidal Love number $k_2$ \cite{Hinderer08,Hinderer09,Hinderer10} from which $\lambda$ can be obtained as
\begin{equation}
\lambda=\frac{2}{3}k_2R^5 \ ,
\label{eq:tidal}
\end{equation}
with $R$ being the radius of the star. It is useful to define the so-called dimensionless tidal deformability $\Lambda$
\begin{equation}
\Lambda=\frac{\lambda}{M^5}=\frac{2}{3}\frac{k_2}{C^5} \ ,
\end{equation}
where $M$ is the mass of the star and, in the rightmost expression, we have introduced the star's compactness parameter $C=M/R$.}

\section{Neutron stars: EoS and Structure}
\label{sec:results}

\begin{figure*}[t!]
\begin{center}
\includegraphics[width=0.49\textwidth]{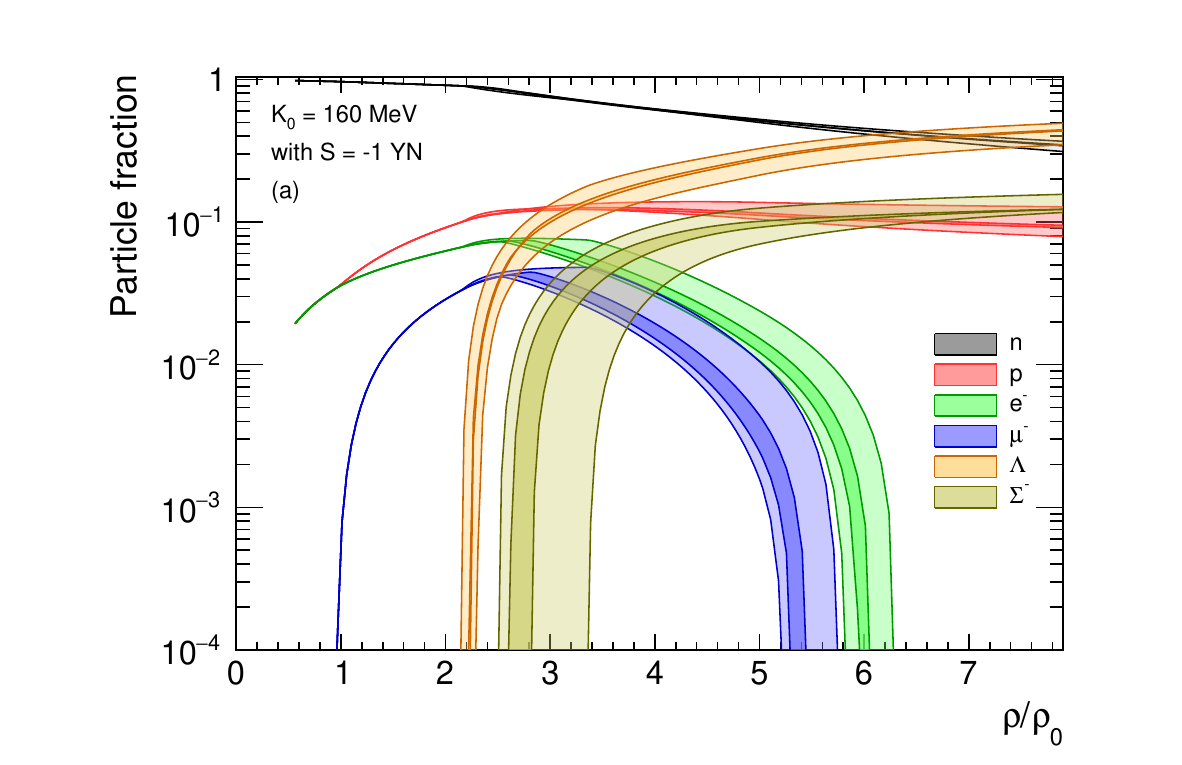}
\includegraphics[width=0.49\textwidth]{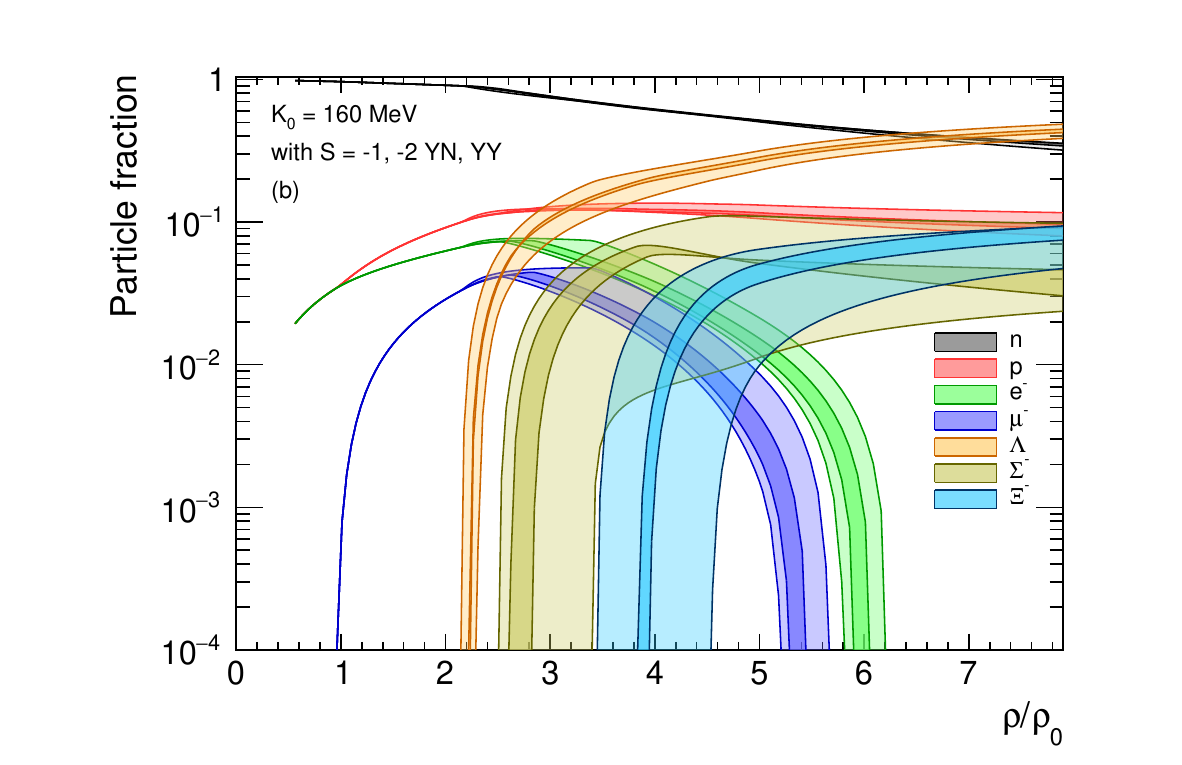}\\
\includegraphics[width=0.49\textwidth]{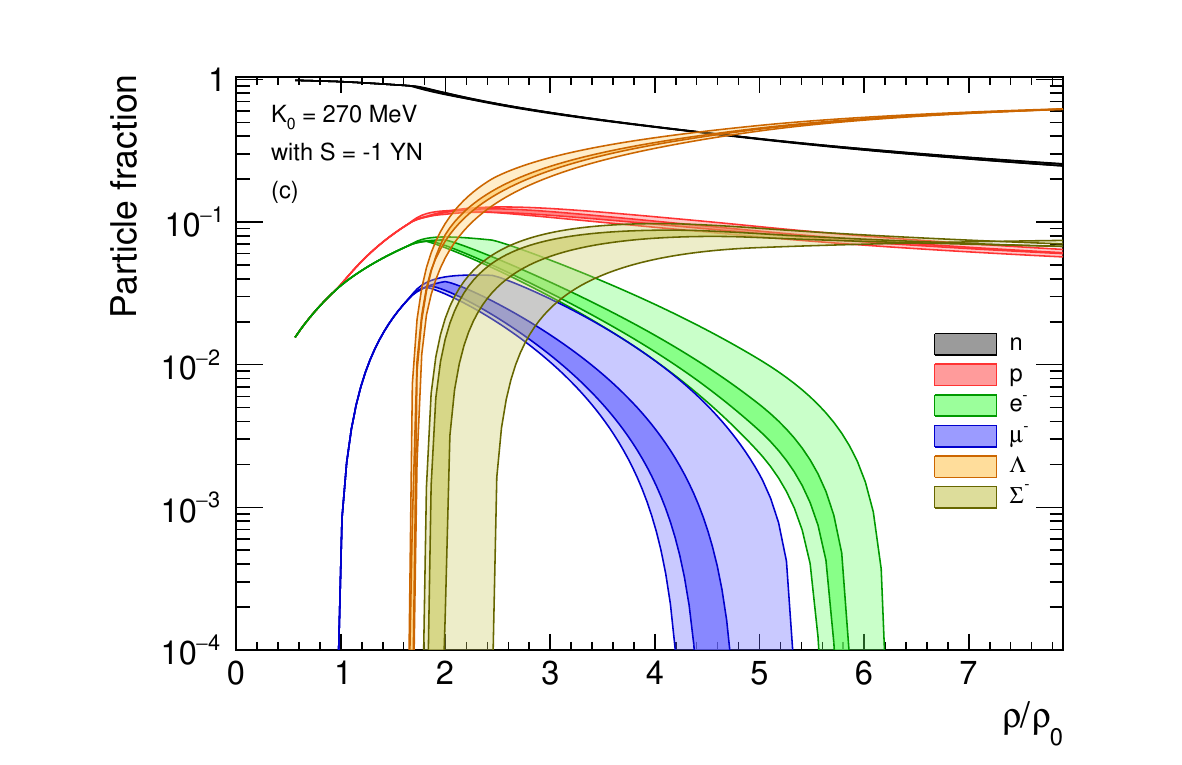}
\includegraphics[width=0.49\textwidth]{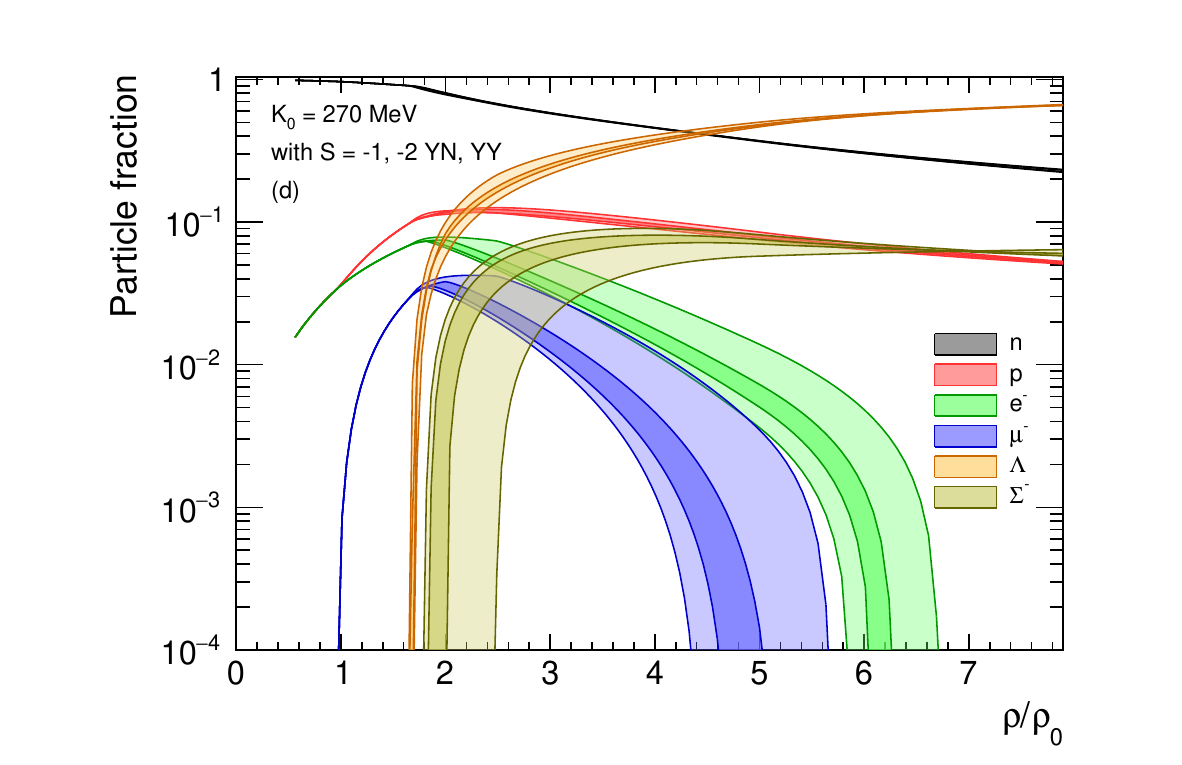}
\caption{(Color online). Composition of $\beta$-stable neutron star matter including nucleons, $\Lambda$, $\Sigma^-$ and $\Xi^-$ baryons. For the $S=-1$ YN interaction we use the NLO19 potential \cite{Haidenbauer00} and
variants established in Ref.~\cite{Mihaylov24}, while 
for the $S=-2$, the  $\Lambda\Lambda$ and $\Xi$N 
potentials from the HAL QCD Collaboration are used,
see text in Sect.~\ref{sec:results}.
The upper (lower) panels (a) and (b) ((c) and (d)) show the results when a soft (stiff) nucleonic EoS with $K_0 = 160$ MeV ($K_0 = 270$ MeV) is considered. Left (right) panels (a) and (c) ((b) and (d)) show the composition assuming only $S=-1$ YN interactions (assuming also $\Lambda\Lambda$ and
$\Xi$N  interactions).
The narrow darker colored bands represent the uncertainty from the femtoscopic 
determination of the $\Lambda p$ scattering lengths~\cite{Mihaylov24}. The wider faint bands show the overall theoretical uncertainty, estimated here from the residual cutoff dependence of the NLO19 interaction.}
\label{fig:fig1}
\end{center}
\end{figure*}

We start this section by showing in Fig.~\ref{fig:fig1} the composition of $\beta$-stable neutron star matter as a function
of the density $\rho$, expressed in units of $\rho_0$. The upper ((a), (b)) and lower ((c), (d)) panels show, respectively, the content of matter when a soft ($K_0=160$ MeV) or a stiffer ($K_0=270$ MeV) nucleonic EoS is considered. 
The particle fractions when only $S=-1$ YN interactions are included in the calculation are displayed in the left ((a), (c)) panels of the figure, whereas those obtained when considering additionally the $S=-2$ $\Lambda\Lambda$ and $\Xi$N interactions are shown in the right ((b), (d)) ones. The different colored bands reflect the uncertainties in the YN and YY interactions. A detailed discussion on how these uncertainties are effectively propagated to the composition, EoS and structure of neutron stars is given at the end of the section. Looking at the left and right panels we observe that, for a fixed nucleonic EoS, the onset density of the $\Lambda$ is the same. This is due to the fact that, up to that density, matter is composed only of nucleons and leptons and, therefore, the $S=-2$ interactions play no role. 
We note also that the density at which the $\Sigma^-$ appears is 
likewise not affected by these interactions simply because the $\Lambda\Sigma^-$ and $\Sigma^-\Sigma^-$ channels have not been included in the calculations. Comparing the upper ((a)-(b)) and lower panels ((c)-(d)) we see that the use of a stiffer nucleonic EoS leads to an earlier onset and a slightly larger population of $\Lambda$ and $\Sigma^-$ hyperons. Additionally, we should notice that whereas the $\Xi^-$ hyperon appears around 4$\rho_0$ when a softer nucleonic EoS is considered, it does not appear (or it appears at very large densities) when using a stiffer one.

\begin{figure*}[t!]
\begin{center}
\includegraphics[width=\columnwidth]{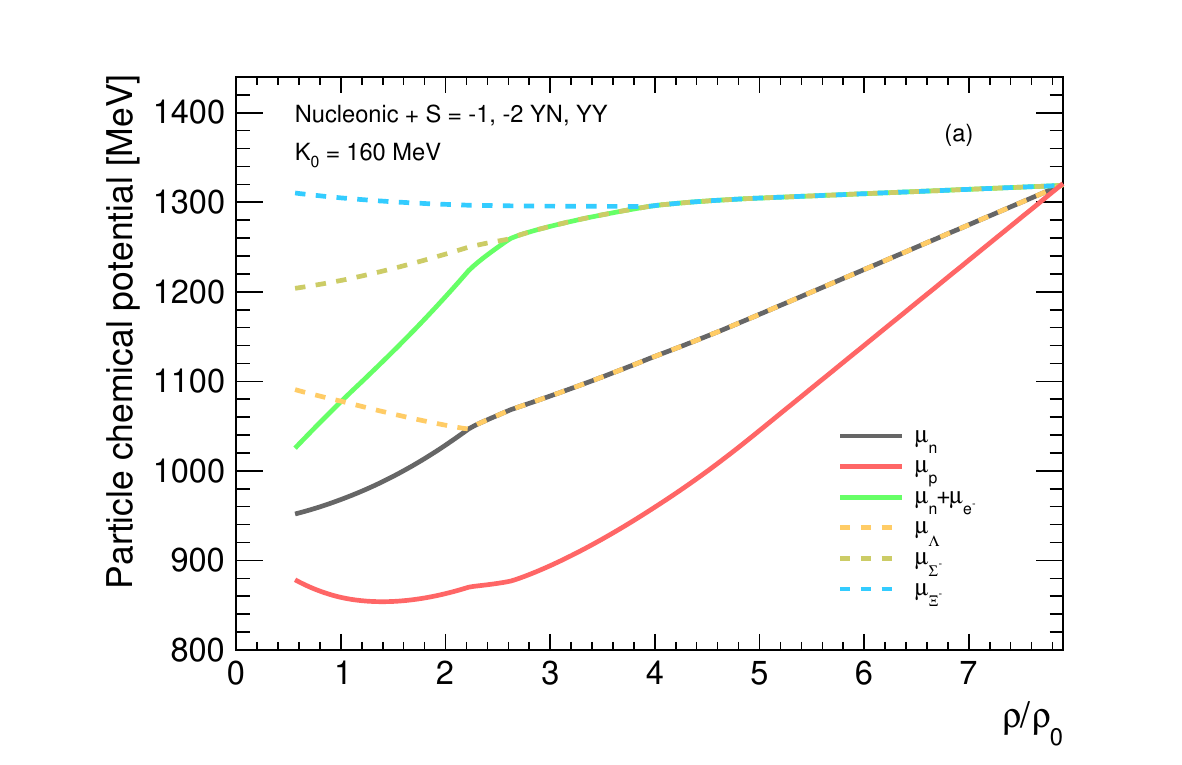}
\includegraphics[width=\columnwidth]{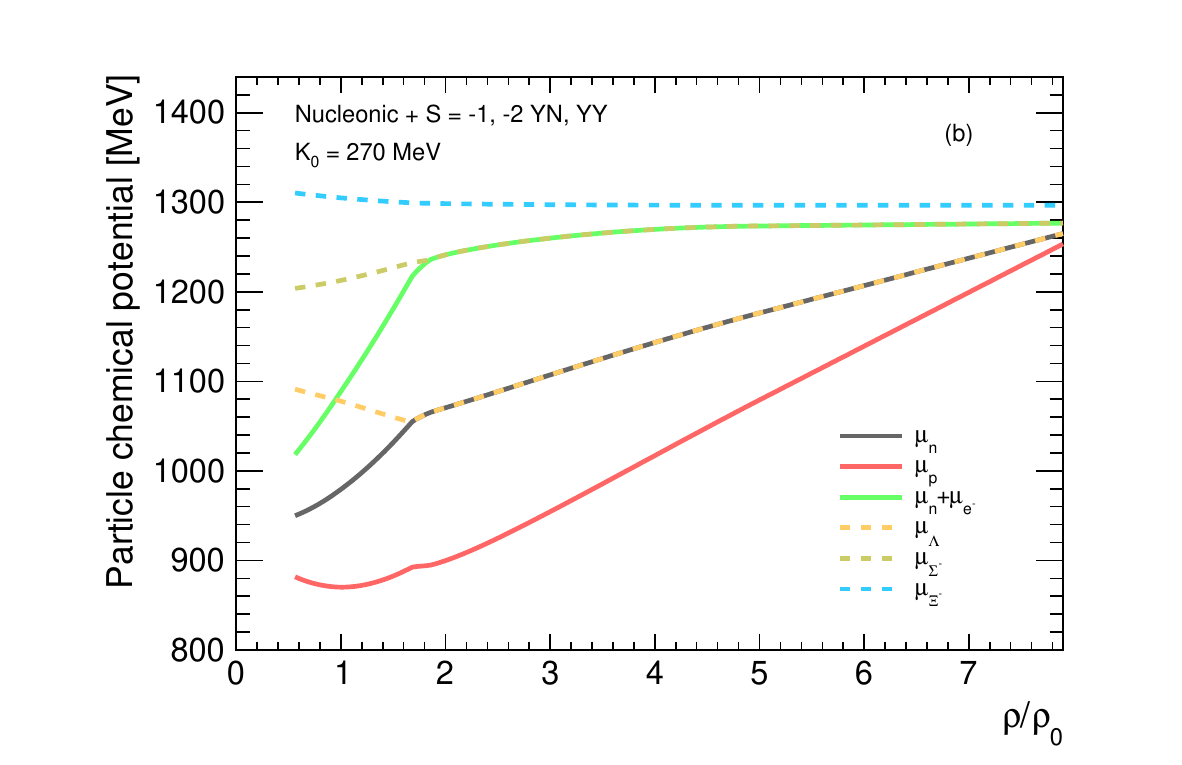}
\caption{(Color online). Particle chemical potentials as a function of the density (expressed in units of $\rho_0$) in $\beta$-stable neutron star matter including nucleons, $\Lambda$, $\Sigma^-$ and $\Xi^-$ baryons when a soft nucleonic EoS with $K_0=160$ MeV is considered (panel (a)) and when a scenario with a stiff nucleonic EoS with  $K_0=270$ MeV is assumed (panel (b)). Results are shown using the chiral YN potential NLO19 with a cutoff of 600 MeV, and 
the $\Lambda\Lambda$  and $\Xi$N HAL QCD potentials for the sets of parameters given in Ref.\ \cite{Sasaki20} for t/a=12.
}
\label{fig:figChemPot}
\end{center}
\end{figure*}

To understand better all these features, we show in Fig.~\ref{fig:figChemPot} the density dependence of the chemical potentials, $\mu_i$, of the different baryon species obtained for the two nucleonic EoS together with the YN
potential NLO19 with a 600 MeV cutoff, and the $\Lambda\Lambda$ and $\Xi$N 
HAL QCD potentials for the sets of parameters given in Tabs.\ 2, 3 and 4 of
Ref.\ \cite{Sasaki20} for t/a=12. Similar qualitative results have been obtained using other values of the cutoff and the parameters of the HAL QCD potentials. The earlier onset of the $\Lambda$ and $\Sigma^-$ hyperons when using a stiffer nucleonic EoS  is now easy to understand since in this case the neutron (black line) and neutron plus electron (green line) chemical potentials increase more rapidly with the density (right panel), ensuring the chemical equilibrium condition for the appearance of these two hyperons is fulfilled at lower densities. Conversely, the softer nucleonic EoS scenario (left panel) leads to a later onset of such hyperons.
The $\Xi^-$ hyperon, which appears for the softer nucleonic EoS at around 4$\rho_0$, does not appear below $\sim$ 8$\rho_0$ for the stiffer one. The reason is that in this case, due to the earlier appearance of the $\Sigma^-$, both the neutron and electron chemical potentials are smaller and, therefore, the $\Xi^-$ onset condition is only fulfilled at very large densities ($\gtrsim$ 9$\rho_0$).

\begin{figure*}[t!]
\begin{center}
\includegraphics[width=\columnwidth]{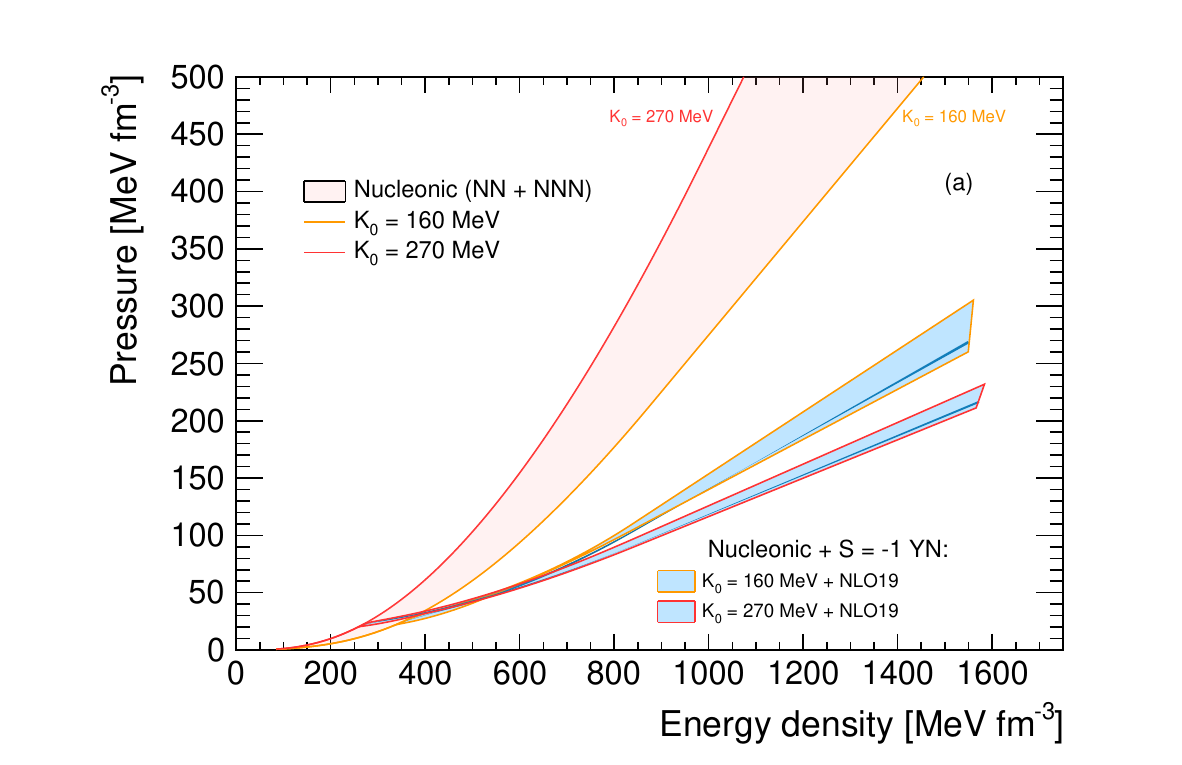}
\includegraphics[width=\columnwidth]{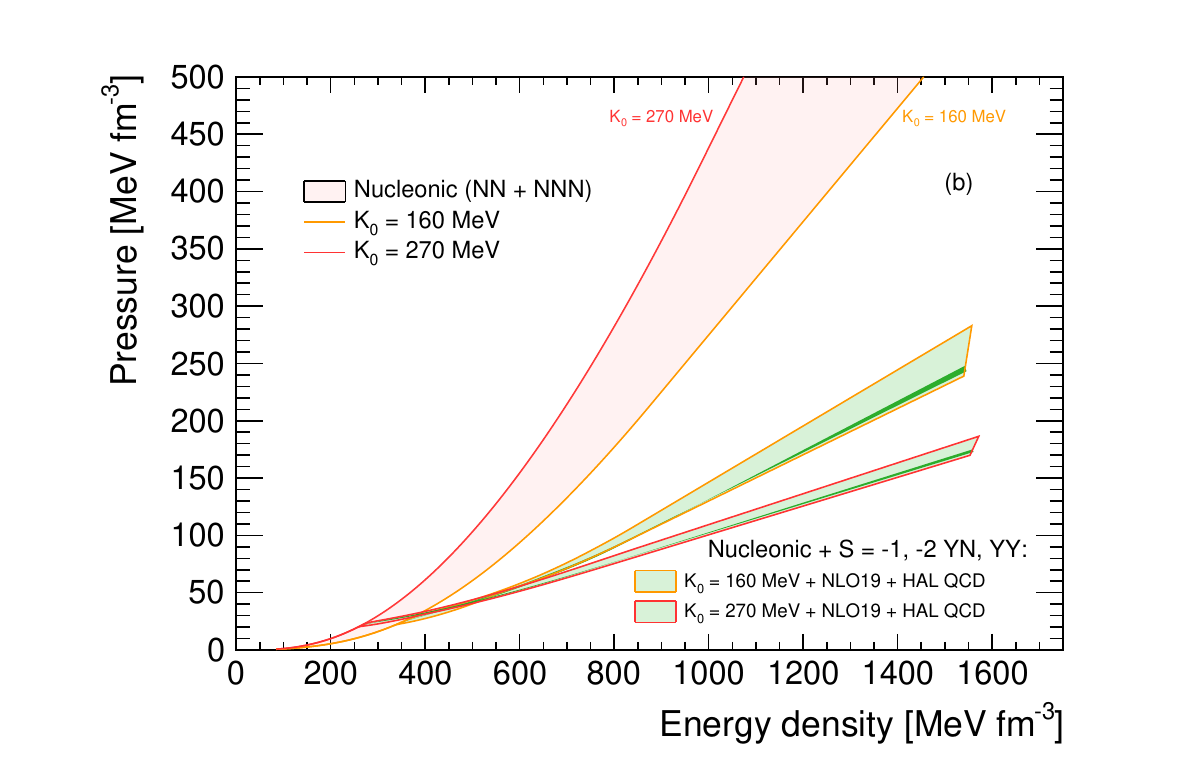}\\
\includegraphics[width=\columnwidth]{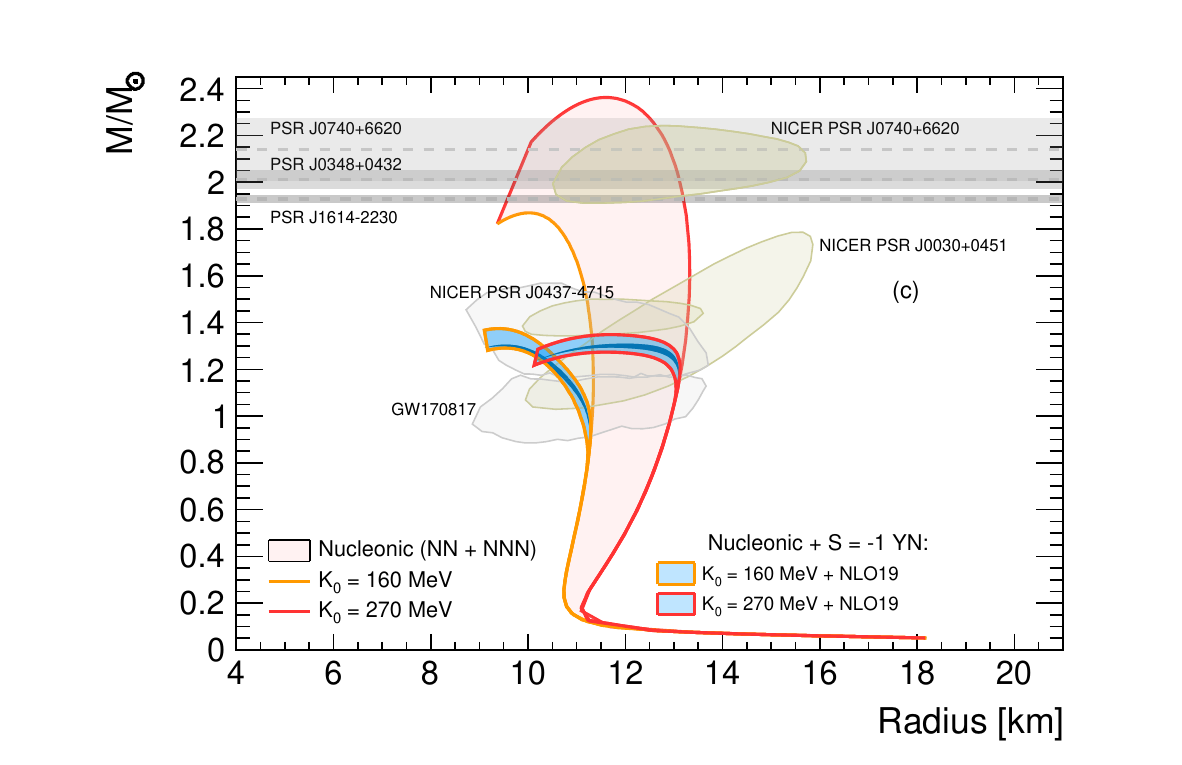}
\includegraphics[width=\columnwidth]{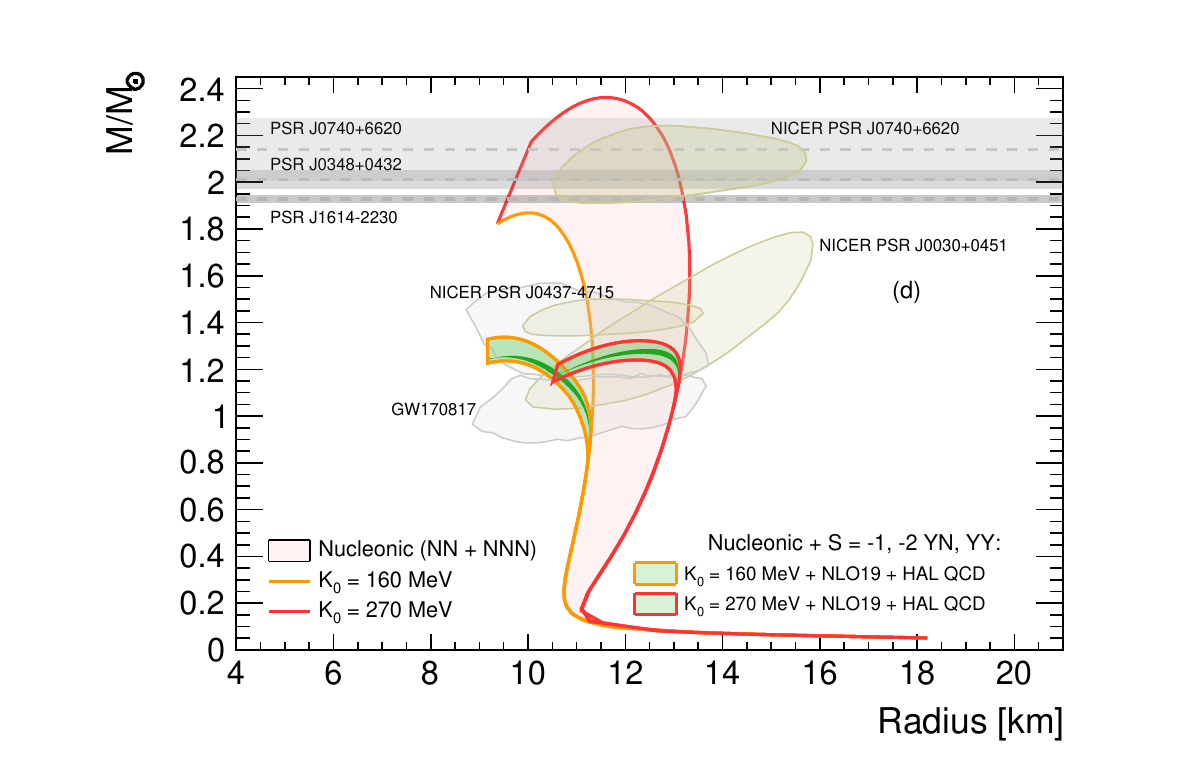}\\
\includegraphics[width=\columnwidth]{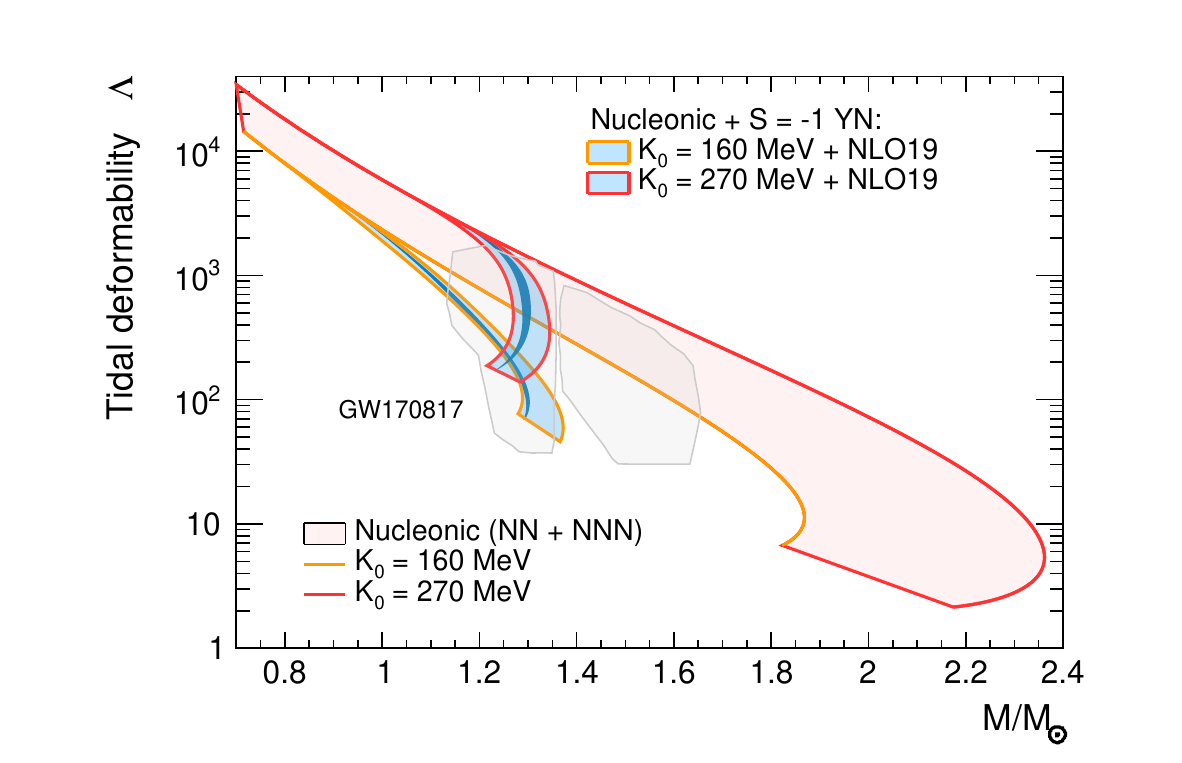}
\includegraphics[width=\columnwidth]{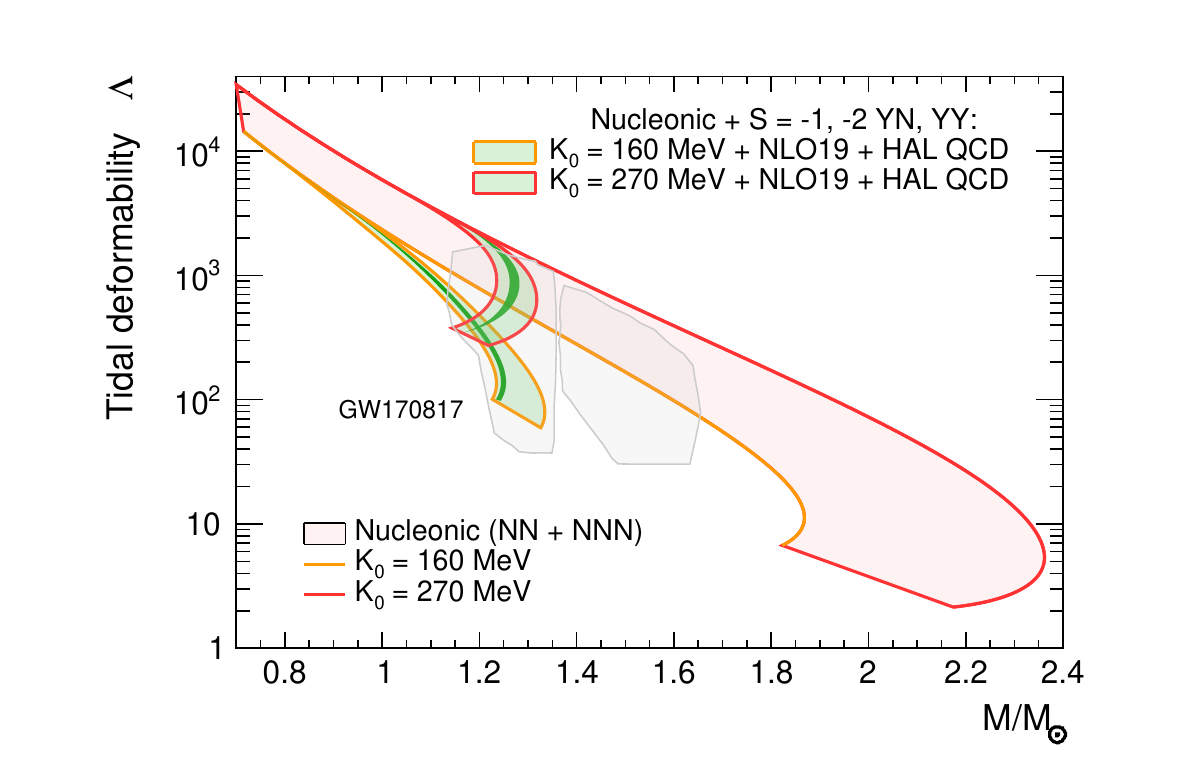}
\caption{(Color online).
Panels (a) and (b): EoS of $\beta$-stable neutron star matter 
including nucleons, $\Lambda$, $\Sigma^-$ and $\Xi^-$ baryons.
The EoS of nucleonic matter in $\beta$-equilibrium 
predicted for $K_0=160$ MeV (orange line) and $K_0=270$ MeV 
(red line) are shown for comparison. For the $S=-1$ YN interaction we use the NLO19 potential \cite{Haidenbauer00} and
variants established in Ref.~\cite{Mihaylov24}, while
for the interactions in the $S=-2$ channels $\Lambda\Lambda$ and $\Xi$N the HAL QCD potentials described in
Sect.~\ref{sec:results} are assumed.
Panels (c) and (d): mass versus radius plots for the corresponding EoS shown in the upper panels. Recent observed masses of the millisecond pulsars 
PSR J1614-2230~\cite{Demorest10}, PSR J0348+0432~\cite{Antoniadis13} and PSR J0740+6620~\cite{Cromartie20} with their  corresponding error bars are also shown, as well as constraints from the gravitational wave event GW170817 \cite{Abbot17}, the NICER mission \cite{Grendeau17,Riley19,Riley21,Vinciguerra24,Salmi24}. The reported astrophysical constraints for the GW170817 and NICER data are provided with 2$\sigma$ uncertainty, while the Shapiro-delay measurements for the heavy pulsars are reported with 1$\sigma$ uncertainty. {Panels (e) and (d): dimensionless tidal deformability $\Lambda$ as a function of the gravitational mass for the corresponding EoS shown in the upper panels. The observational constraint inferred in Ref.\ \cite{Kumar23} from GW170817 are shown with 2$\sigma$ uncertainty.}}
\label{fig:fig2}
\end{center}
\end{figure*}

The neutron star EoS and the mass-radius relation,
 corresponding to the particle compositions of the different cases just described, are respectively shown in the upper ((a), (b)) and middle ((c), (d)) panels of Fig.~\ref{fig:fig2}. 
{The dimensionless tidal deformability $\Lambda$ is displayed in panels (e) and (f).} Results assuming only $S=-1$ YN potentials are displayed on the left ((a), (c), (e)) panels whereas those including additionally the $\Lambda\Lambda$ and $\Xi$N interactions in the $S=-2$ channels are presented on the right ((b), (d), (f)) ones. 
In the middle panels (c) and (d), we also show the recently observed masses of three heavy millisecond pulsars PSR J1614-2230 ($1.928 \pm 0.017 M_\odot$)~\cite{Demorest10}, PSR J0348+0432 ($2.01 \pm 0.04 M_\odot$)~\cite{Antoniadis13} and PSR J0740+6620 ($2.14^{+0.10}_{-0.09} M_\odot$)~\cite{Cromartie20} with their corresponding uncertainty bars, as well as constraints from the neutron star merger gravitational wave event GW170817~\cite{Abbot17}, and from the NICER mission \cite{Grendeau17,Riley19,Riley21,Vinciguerra24,Salmi24}. {Constraints on the tidal deformability inferred from GW170817 in Ref.\ \cite{Kumar23} are also reported in panels (e) and (f) of the figure.} 

Let us focus first on the EoS. As it is expected
\cite{Chatterjee16,Vidana18}, the appearance of hyperons leads to a softening of the EoS that is enhanced when the nucleonic contribution to the EoS is stiffer. This is because, in this case, the onset of the $\Lambda$ and $\Sigma^-$ hyperons is shifted to lower densities and their population is larger (see Fig.~\ref{fig:fig1}), as discussed before. A more quantitative estimate of such a softening effect can be obtained by comparing for instance the small dark blue bands in panel (a) or the small green ones in panel (b) of Fig.~\ref{fig:fig2}
for the soft and stiff nucleonic scenario. Particularly, in the high energy density region, one can observe a sizable reduction in the pressure of the order of about $20-25\%$ when a stiff nucleonic EoS is used. As it is seen also in panel (b), an additional softening of the EoS is obtained when we take into account the effect of the
$\Lambda\Lambda$ and $\Xi$N interactions, due to their attractive character.

We now turn to describe the effects of the different EoS on the mass-radius relation of neutron stars. Results are shown in panels (c) and (d) of Fig.~\ref{fig:fig2}. We note first that the maximum mass of a neutron star with $S=-1$ and $S=-2$ hyperons in its interior is around $1.3-1.4$ $M_\odot$, as shown in panels (c) and (d) with the green and blue bands, respectively. We note also that such values seem to be quite insensitive to the nucleonic part of the EoS, washing out the rather large uncertainty in the maximum mass (from $\approx 2.3$ M$_\odot$ to $1.8$ M$_\odot$) achieved with a nucleons-only scenario. This insensitivity to the nucleonic EoS was already observed in Refs.\ \cite{Schulze06,Schulze11} where the Nijmegen YN and YY meson-exchange interaction models were used in combination with different NN potentials. As it was argued in these two works the reason for this should be traced to several strong compensation mechanisms caused by the appearance of hyperons that always leads to a soft EoS keeping the value of the maximum mass within $\sim 0.05\, M_\odot$. Note, however, that there still remain characteristic differences between the mass-radius relations that would eventually allow one to determine the EoS from observational constraints. In particular, there are significant differences for the corresponding radii achieved within the different EoS assumptions, which are linked to the maximum central baryon density. The overall stiffest EoS scenario with hyperons, achieved for the model with low nuclear incompressibility  $K_0=160$ MeV, typically leads to radii of approximately 10 km, while larger values ($\approx 12$ km) are reached when the nuclear incompressibility is as well larger.
Concerning the agreement with the reported measured masses and radii, we observe that the heavy pulsars data can be only reproduced by pure nucleonic EoSs.
On the other hand, the gravitational wave constraints from GW170817, along with the NICER mass and radius measurements inferred for PSR J0030+0451, can be overall reproduced with the softer EoS with strange and multistrange hyperons. We note, however, that the latter shows a slight tension with the most recent NICER constraints on the PSR J0437-4715 millisecond pulsar.

{Let us now analyze the effect of the hyperons on the dimensionless tidal deformability parameter $\Lambda$ shown as a function of the stellar mass in panels (e) and (f) of Fig.\ \ref{fig:fig2} for the different EoS considered together with the constraints inferred in Ref.\ \cite{Kumar23} from the GW170817 event. Note first that for any given value of the stellar mass, the tidal deformability of a neutron star containing hyperons is always smaller than that of a pure nucleonic one. This is simply so because its radius is smaller than that of a pure nucleonic star with the same mass (see panels (c) and (d)). Therefore, its compactness $C$ is larger and, consequently, its tidal deformability is smaller. Note also that while the tidal deformability predicted by the pure nucleonic EoSs is fully compatible with the constraints inferred from GW170817, the  one predicted by the hyperonic ones are in agreement with observation only in the range  $1.1-1.3$ $M_\odot$. However, we recall that none of the hyperonic EoSs considered in this work predicts a maximum mass larger than $1.3-1.4$ $M_\odot$. Therefore, we cannot expect any agreement with observation for masses above these values in the case of hyperonic stars.}

{Finally, we describe how  the uncertainties in the hyperon interactions are effectively propagated to the composition, EoS, mass-radius relation {and tidal deformability} of neutron stars. These uncertainties, shown by the bands in Figs.\ \ref{fig:fig1} and \ref{fig:fig2}, originate almost entirely from the interaction in the $\Lambda$N-$\Sigma$N system}. The bands are constructed in the following way: For each one of the two pure nucleonic EoS, the wider bands are built by performing all possible calculations combining the NLO19 YN interactions with cutoff values 500-650 MeV
(variants II in Tab.~\ref{tab:LN}), with two $\Xi$N HAL QCD potentials corresponding to the two sets of parameters {selected out of the 71 ones employed by the ALICE Collaboration in its analysis of the $\Xi^-p$ correlations made in Ref.\ \cite{ALICE:pXiNature}, and the $\Lambda\Lambda$ and $\Lambda\Lambda-\Xi$N potentials for the set of parameters listed in Tables 2 and 3 of Ref.\ \cite{Sasaki20} corresponding to $t/a={11}$}\footnote{{We have verified that the different parameterizations of the $\Lambda\Lambda$ and $\Lambda\Lambda-\Xi$N potentials from the HAL QCD Collaboration lead to almost identical results within the precision of the present analysis. Thus, in this work only the set of parameters for $t/a=11$ has been considered.}}. The upper and lower extremes of the wider bands are then defined by the combination which gives, respectively, the stiffer and softer EoS. Similarly, the narrow bands are constructed by using the variants of the NLO19 YN interaction with cutoff 600 MeV, with LECs tuned to femtoscopic $\Lambda p$ data
(variants I in Tab.~\ref{tab:LN}), combined with the same $\Xi$N, 
$\Lambda\Lambda$ and $\Lambda\Lambda-\Xi$N potentials. The extremes of the narrow bands are defined as those of the wider ones. 

Not unexpectedly, the theoretical uncertainty due to the underlying $\Lambda$N-$\Sigma$N interaction dominates the results. In fact, in analogous studies of (purely) nuclear matter 
properties based on NN interactions derived within $\chi$EFT the theoretical uncertainty was found to be likewise large~\cite{Hu:2016nkw,Sammarruca:2014zia}. Despite of much higher orders in the chiral expansion (N$^4$LO) a reasonable convergence was only observed up to about  twice the saturation density of nuclear matter~\cite{Hu:2016nkw}.
Clearly, for improvements, among other things, a consistent inclusion of corresponding many-body forces is required \cite{Sammarruca:2014zia}. We want to mention that there is an additional source of uncertainty due to the fact that the $\Lambda N$ interaction in $P$- and higher partial waves is not constrained by empirical information. Judging from results in
the literature, for example those in Ref.~\cite{Petschauer:2015nea}, it could be in the order of $\pm 3$~MeV in the $\Lambda$ single-particle potential at nuclear matter saturation density. We do not consider this uncertainty in the present calculation.

\section{Summary and Conclusions}
\label{sec:conclusions}

In this work we have employed the ab-initio microscopic Brueckner--Hartree--Fock theory extended to the strange baryon sector to construct the EoS of hypernuclear matter under $\beta$-equilibrium conditions, relevant for the study of neutron stars. To such end, we have used a chiral hyperon-nucleon interaction of the J\"{u}lich--Bonn group tuned to femtoscopic $\Lambda p$ data of the ALICE Collaboration, and $\Lambda\Lambda$ and $\Xi$N interactions determined from lattice QCD calculations by the HAL QCD Collaboration that reproduce the femtoscopic $\Lambda\Lambda$ and $\Xi^-p$ data. To describe the pure nucleonic part of the EoS we have used the well-known Argonne V18 nucleon-nucleon potential. To account for the effect of three-nucleon forces we have added a phenomenological term to the total energy density of the system with parameters fitted to reproduce simultaneously the binding energy of symmetric nuclear matter at saturation and the nuclear incompressibility $K_0$.

We have put a special focus on the interaction uncertainties and how they are effectively propagated to the composition, EoS and mass-radius relation of neutron stars. In particular, to explore the uncertainty on the neutron star properties associated to the softness/stiffness
of the pure nucleonic part we have considered two extreme values of $K_0$ compatible with current heavy-ion experimental data. Regarding  
hyperons, we have considered the uncertainty due to the experimental error of the femtoscopic data used to fix the chiral hyperon-nucleon interaction and the theoretical uncertainty estimated from the residual cut-off dependence of this interaction, as well as the uncertainties on the $\Lambda\Lambda$ and $\Xi$N interactions inferred from the analysis of correlation measurements. We have employed two specific sets of parameterizations of the $\Xi$N potential from lattice QCD~\cite{Sasaki:2020}, out of the 71 explored in the ALICE analysis of $\Xi^-p$ correlations, that correspond, respectively, to the most and less attractive $\Xi$N compatible with femtoscopic data. We have seen that the different parametrizations of the HAL QCD Collaboration for the $\Lambda\Lambda$ potential and the transition $\Lambda\Lambda-\Xi$N lead to identical results within the precision of the present analysis. Therefore, we have considered only one of the parameters sets listed in Tables~2 and 3 of Ref.\ \cite{Sasaki20}. 

Our results have shown, in agreement with previous works, that the appearance of hyperons in the interior of neutrons of neutron stars leads to a strong softening of the EoS and to a reduction of the maximum mass to values in the range $1.3-1.4\, M_\odot$,  incompatible with current observations. In particular, the sizable range of maximum neutron star masses achieved within a pure nucleonic EoS, constrained to the current incompressibility limits, is significantly reduced once hyperons appear, and the final maximum mass of a hyperonic star results quite independent of the nucleonic component of the EoS, as it was already observed in Refs.\ \cite{Schulze06,Schulze11}. Therefore, the hyperon puzzle remains still an open issue if only two-body hyperon-nucleon and hyperon-hyperon interactions are considered. Its solution requires a mechanism or mechanisms that could provide the additional repulsion that makes the EoS stiffer and, therefore, the maximum mass compatible with observations. 
One of such mechanisms could be the inclusion of hyperonic three-body forces 
\cite{Petschauer:2015elq,Haidenbauer:2016vfq,Kohno:2018gby,Logoteta:2019utx,Gerstung:2020ktv,Tong:2024egi} or, for instance, to consider a more exotic scenario where light QCD axions could appear in neutron stars \cite{Balkin:2022qer}. Such possible mechanisms will be analyzed in future works. {Finally, we have found that the tidal deformability of neutron stars with hyperons are in agreement with the observational constraints from the GW170817 event in the mass range $1.1-1.3$ $M_\odot$.}

\section*{Acknowledgments}
This work was supported by ORIGINS cluster DFG under Germany’s Excellence Strategy-EXC2094 - 390783311 and the DFG through the Grant SFB 1258 ``Neutrinos and Dark Matter in Astro and Particle Physics”. I.V. acknowledges the Excellent Cluster ORIGINS cluster DFG under Germany’s Excellence Strategy-EXC2094 - 390783311 for its support during his visit at TUM in June and July 2024.
V. M. S. was supported by the Deutsche Forschungsgemeinschaft (DFG) through the grant MA $8660/1-1$. 



\end{document}